\documentclass[12pt]{article}
\usepackage{graphicx}
\newcommand{\zobs}{\zeta_{\rm obs}}
\newcommand{\phizet}{(\phi,\zobs)}
\newcommand{\rovc}{r_{\rm ovc}}
\newcommand{\rmax}{r_{\rm max}}
\newcommand{\rhomax}{\rho_{\rm max}}
\newcommand{\rlc}{R_{\rm lc}}
\newcommand\bm{\boldmath}
\newcommand\ubm{\unboldmath\ }

\title{Polarization of pulsar radiation in the two-pole caustic model
and in the outer gap model\footnote{Poster presented at the X-Ray
Polarimetry Workshop, SLAC, Stanford, California, $9-11$ February
2004.}}
\author{J.~Dyks$^{1,2}$, A.~K.~Harding$^1$,
and B.~Rudak$^2$}

\footnotetext[1]{NASA Goddard Space Flight Center, Greenbelt, MD, USA}
\footnotetext[2]{Nicolaus Copernicus Astronomical Center, Toru\'n,
Poland}

\begin{document}

\maketitle

\def \rns{R_{\scriptscriptstyle NS}}
\def \delp {\Delta^{\rm peak}}
\def\la{\hbox{\hspace{1.5mm}}\raise2pt
       \vbox{\hbox{$<$}}\lower2pt
       \vbox{\moveleft9.0pt\hbox{$\sim$ }}\hbox{\hskip 0.05mm}}
\def\gap{\hbox{\hspace{1.5mm}}\raise2pt
       \vbox{\hbox{$>$}}\lower2pt
       \vbox{\moveleft9.0pt\hbox{$\sim$ }}\hbox{\hskip 0.05mm}}
\def\ga_Large{\hbox{\hspace{1.5mm}}\raise3pt
       \vbox{\hbox{$>$}}\lower3pt
       \vbox{\moveleft13.0pt\hbox{$\sim$ }}\hbox{\hskip 0.05mm}}


\section*{Abstract}
We present linear polarization characteristics of pulsar radiation as
predicted by two models of high-energy radiation from extended
regions in the outer magnetosphere: the recently proposed two-pole
caustic (TPC) model (Dyks \& Rudak 2003), and the outer gap (OG) 
model (Romani \& Yadigaroglu 1995 -- RY95; Cheng, Ruderman \& Zhang 2000).

\section{Two-pole caustic model vs outer gap model}

The two-pole caustic model differs from the outer gap model in that
its emission region extends below the null charge surface toward the
polar cap (Fig.~1). The TPC model discards emission from regions close
to the light cylinder (ie.~from $\rho \gap 0.8\rlc$ 
where $\rho$ is the distance
from the rotation axis and $\rlc$ is the radius of the light cylinder),
because the dynamics of electrons and the geometry of the magnetic field
is not known there whereas the OG model includes this emission near the
light cylinder. The TPC model assumes that the bulk of photon
emission occurs on the last open field lines ($\rovc=1$) whereas the OG model
employs $\rovc = 0.8-0.9$, where $\rovc$ is the ``open volume coordinate"
corresponding to the location of the gamma-ray emitting magnetic field 
lines
at the star surface ($\rovc = 1$ -- rim of the polar cap; $\rovc \sim 0$
-- central parts of the polar cap).
Due to these differences in geometry of the emission region, the TPC
model uses the double-pole interpretation of the widely separated 
double peaks of some gamma-ray pulsars (Crab, Vela, Geminga, B1951$+$32,
B1046$-$58). The OG model offers the single-pole interpretation.
Strong differences between the emission patterns of the two models,
and between predicted lightcurves are shown in Fig.~2. 

\newpage

\section{Calculation method}

The vacuum dipole with the sweepback effect is used.
Uniform intensity per unit length of the magnetic field lines is
assumed in the corotating frame (CF). The electric field vector of the
received radiation $\vec E_{\rm w}$ is taken along the local acceleration 
vector $\vec a$ at the emission point (RY95 took $\vec E_{\rm w}
\parallel \vec \rho_{\rm curv}$ where $\vec \rho_{\rm curv}$ is the local
curvature radius of the magnetic field lines in the CF). The intrinsic
polarization degree of $80\%$ was assumed. To account for possible
overlap of emissions from different regions of the magnetosphere,
appropriate handling of Stokes parameters $I$, $Q$, $U$ was performed.
The parameters were finally transformed to the position angle $\psi$ and
the polarization degree $P[\%]$.

\section{Polarization data}

So far, ``high-energy" data on pulsar polarization are limited to
the optical data on the Crab pulsar (Jones et al.~1981; Smith et al.~1988; 
Graham-Smith et al.~1996; Romani et al.~2001; Kanbach et
al.~2003) and optical data on B0656$+$14 (Kern et al.~2003). 
The left column of Fig.~3 presents the lightcurve (top
panel),
the position angle (PA) curve (middle panel), and the degree of
polarization (bottom panel) for the Crab pulsar observed with the OPTIMA
instrument (Kanbach et al.~2003; poster at this workshop). 
Two fast swings of position angle
and very low polarization degree at both peaks are noticeable.
Beyond the peaks the signal is apparently dominated by a component
with fixed position angle, so far of unrecognized origin.
The right column shows the data with the constant component subtracted
(Kellner 2002). A similar anticorrelation between the total flux and the
polarization degree has also been observed in the optical for
B0656$+$14 (Kern et al.~2003).

\section{Modelled polarization properties of pulsars}

Radiation characteristics calculated for the TPC model 
with the dipole inclination $\alpha=70^\circ$ are presented
in Fig.~4, which consists of nine three-panel frames for nine different
viewing angles $\zobs$ measured from the rotation axis (their values are
displayed in the upper right corner of each frame). 
Each frame shows the lightcurve (top), the position angle
$\psi$ (middle), and the polarization degree (bottom).
Photon emission assumed in this calculation was dominated by, but
not limited to, the last open field lines. 
Emission from neighboring magnetic field lines (with different $\rovc$)
was weighted by $\exp[0.5(\rovc - 1)^2/\sigma^2]$ with $\sigma = 0.025$, 
ie.~a Gaussian profile
centered at the polar cap rim was assumed.  

For most viewing angles the double-peaked lightcurves can be discerned.
Associated with the peaks are fast changes of $\psi$ and minima in $P[\%]$
which are similar to those observed in the Crab. 
The fast change of $\psi$ at the leading peak near $\phi \simeq 0.1$
is due to caustic effects
and is faster on the trailing side of the peak, than on the leading
side, in agreement with the Crab data. This swing
is a very stable feature whereas the swing at the trailing peak
is very sensitive to model parameters. 
The minima in $P[\%]$
have the form of ``double dips" -- the leading dip in each pair is 
due to a combination of caustic effects and overlapping emission from 
the neighboring
magnetic field lines with slightly different $\rovc$. The dips which lag the
maxima in intensity arise because of superposition of radiation patterns
from opposite magnetic hemispheres. For more details see Dyks, Harding \&
Rudak 2004. 

In Fig.~5 results for the OG model are shown with the same layout as
before. Again, $\alpha=70^\circ$ and the Gaussian intensity profile,
this time centered at $\rovc=0.85$, were assumed.
In spite of the similar (ie.~caustic) nature of the profiles' peaks, 
the position angle curves and polarization degree predicted by the OG
model are completely different than those observed for the Crab.
Apparently, this demonstrates the sensitivity of the modelled results
to the parameter $\rovc$.
The difficulty of the OG model in reproducing the Crab's data
persists for other model parameters.
We find that the OG model is not able to reproduce the Crab optical data even
qualitatively.

\section{Comparison with the outer gap results of RY95}

The polarization characteristics shown in Fig.~5 differ significantly
from those presented by RY95 in their fig.~5. 
We argue that our results are superior to those of RY95 because their
calculation method was flawed at least in two aspects.
First, to take into account contributions of emission 
from different regions of the magnetosphere at the same rotational phase,
they just ``averaged the position angles" (see Section 3 in RY95) 
instead of appropriate
handling of the Stokes parameters. Second, they neglected to include
the acceleration of an electron due to the corotation of the
magnetosphere. The latter effect considerably affects the position angle
already at $r \sim 0.1\rlc$, not to mention the vicinity of the light
cylinder (cf.~fig.~3 in Hibschman \& Arons 2001).

Our attempts to reproduce RY95' results for the Crab pulsar (fig.~5 in
RY95) fail already at the level of the lightcurve.
Fig.~6 presents the radiation pattern on the $(\phi, \zobs)$ plane,
the lightcurve, the position angle, and the degree of polarization
(top to bottom) for the same parameters as in RY95
(ie.~$\alpha=80^\circ$, $\zobs=62^\circ$).
We obtain a different lightcurve even though we exactly followed their
prescription for emissivity along magnetic field lines ($F\propto
2s-s^2$, with a gaussian decline above $s=1$ with $\sigma=0.5$, see RY95
for details) as well as the same formula for footprints of the
gamma-ray bright magnetic field lines
($w=0.02(70^\circ/\alpha)(r_c/r_{c{\rm , min}})$, where $w \approx 1 - \rovc$, 
for definitions see RY95).
The location of these footprints 
on the polar cap is shown in Fig.~7 for selected magnetic field lines.

Although we sampled a very large variety of model parameters 
(including different prescriptions for the emissivity and for the open
volume coordinate $w$), we were
not able to obtain the lightcurve and the position angle swing similar 
to the one in fig.~5 of RY95 for the same $\alpha$ and $\zobs$. 
All evidence we have gathered, and especially the comparison of our
radiation pattern with the one obtained by RY95 (R.~W.~Romani, private
communication) suggest that their result was obtained for a 
``very specialized" selection of the gamma-ray-bright magnetic field lines 
which differs from the formula given in their paper.
 
\section{Conclusions}

None of the models was able to exactly reproduce the optical polarization
data on the Crab pulsar. The general features, however, (ie.~fast changes 
of position
angle and minima in polarization degree associated with two peaks)
find qualitative explanation within the two-pole caustic model.   

\section{Discussion}

Our calculation method was purely geometrical and neglected
any details of physics of radiation mechanism(s).
The only ``high-energy" aspect of this calculation was
the radially extended emission region (at radio wavelengths the radial
extent is usually assumed to be small). 
Extension of this work to include microscopic physics 
would be of great value (cf.~Epstein 1973; Chen et al.~1996).

Muslimov \& Harding (2004) have recently extended the slot gap acceleration 
to very high altitudes, which makes the geometry of the slot gap model
similar to the TPC model. The accelerating electric field which they
propose may be used to model the physics of emission and to improve
the polarization calculation for the TPC/slot gap model.

\def \bi {\par \noindent \hangindent=0.7cm\hangafter1}
\section*{References:}
\baselineskip = 5mm
\noindent
\bi Chen, K., Chang, H.-K., \& Ho, C. 1996, ApJ, 471, 967
\bi Cheng, K. S., Ruderman, M. A., \& Zhang, L. 2000, ApJ, 537, 964
\bi Dyks, J., \& Rudak, B., 2003, ApJ, 598, 1201
\bi Dyks, J., Harding, A.K., \& Rudak, B. 2004, ApJ, in press
  (astro-ph/0401255)
\bi Epstein, R. I., 1973, ApJ, 183, 593
\bi Graham-Smith, F., Dolan, J. F., Boyd, P. T.,
  Biggs, J. D., Lyne, A. G., et al. 1996, MNRAS, 282, 1354
\bi Hibschman, J. A., \&  Arons, J. 2001, ApJ, 546, 382
\bi Jones, D. H. P., Smith, F. G., \& Wallace, P. T. 1981, MNRAS, 196, 943
\bi Kanbach, G., Kellner, S., Schrey, F., Steinle,
  H., Straubmeier, C., et al. 2003, in Proc.~of the 
  SPIE Meeting on Astronomical
  Telescopes and Instrumentation "Power Telescopes and Instrumentation
  into the New Millenium", eds. Iye, M., \& Moorwood, A. F., 4841, 82
\bi Kellner, S., 2002, Diplomarbeit, Technische Universit\"at M\"unchen
\bi Kern, B., Martin, C., Mazin, B., \& Halpern, J.P. 2003, ApJ, 597,
    1049 
\bi Muslimov, A.G., \& Harding, A.K. 2004, ApJ, in press
\bi Romani, R. W. \& Yadigaroglu, I.-A., 1995, ApJ, 438, 314 (RY95)
\bi Romani, R. W., Miller, A. J., Cabrera, B.,  
  Nam, S. W., \& Martinis, J. M. 2001, ApJ, 563, 221
\bi Smith, F. G., Jones, D. H. P., Dick, J. S. B.,
  \& Pike, C. D. 1988, MNRAS, 233, 305

\begin{figure}[!t]
\centering
\includegraphics[width=0.7\textwidth]{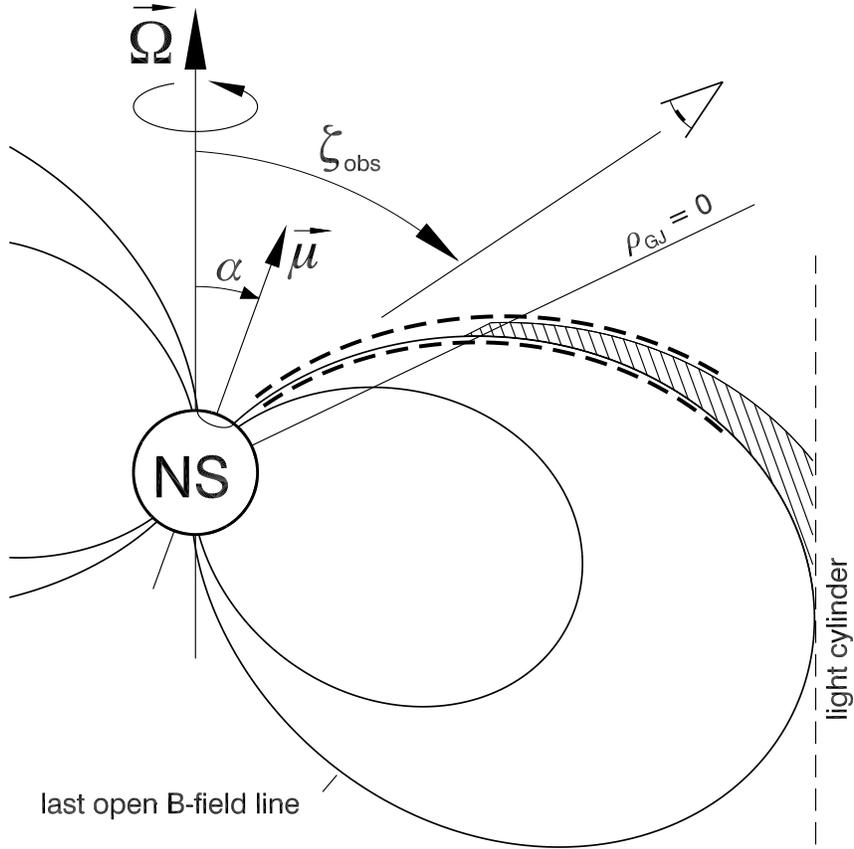}
\caption{Illustration of the two-pole caustic model.
The radiating region (within the dashed lines) is confined to the
surface of the last open field lines, and it extends from the polar cap
to the light cylinder. For comparison, the conventional outer gap
region is shown (shaded area), extending from the surface of the null
space-charge ($\rho_{\rm GJ}=0$, where $\rho_{\rm GJ}\approx
-\vec\Omega\cdot\vec B(2\pi c)^{-1}$ is the Goldreich-Julian charge
density) to the light cylinder.
}
\label{f1}
\end{figure}

\begin{figure}[!t]
\centering
\includegraphics[width=0.6\textwidth]{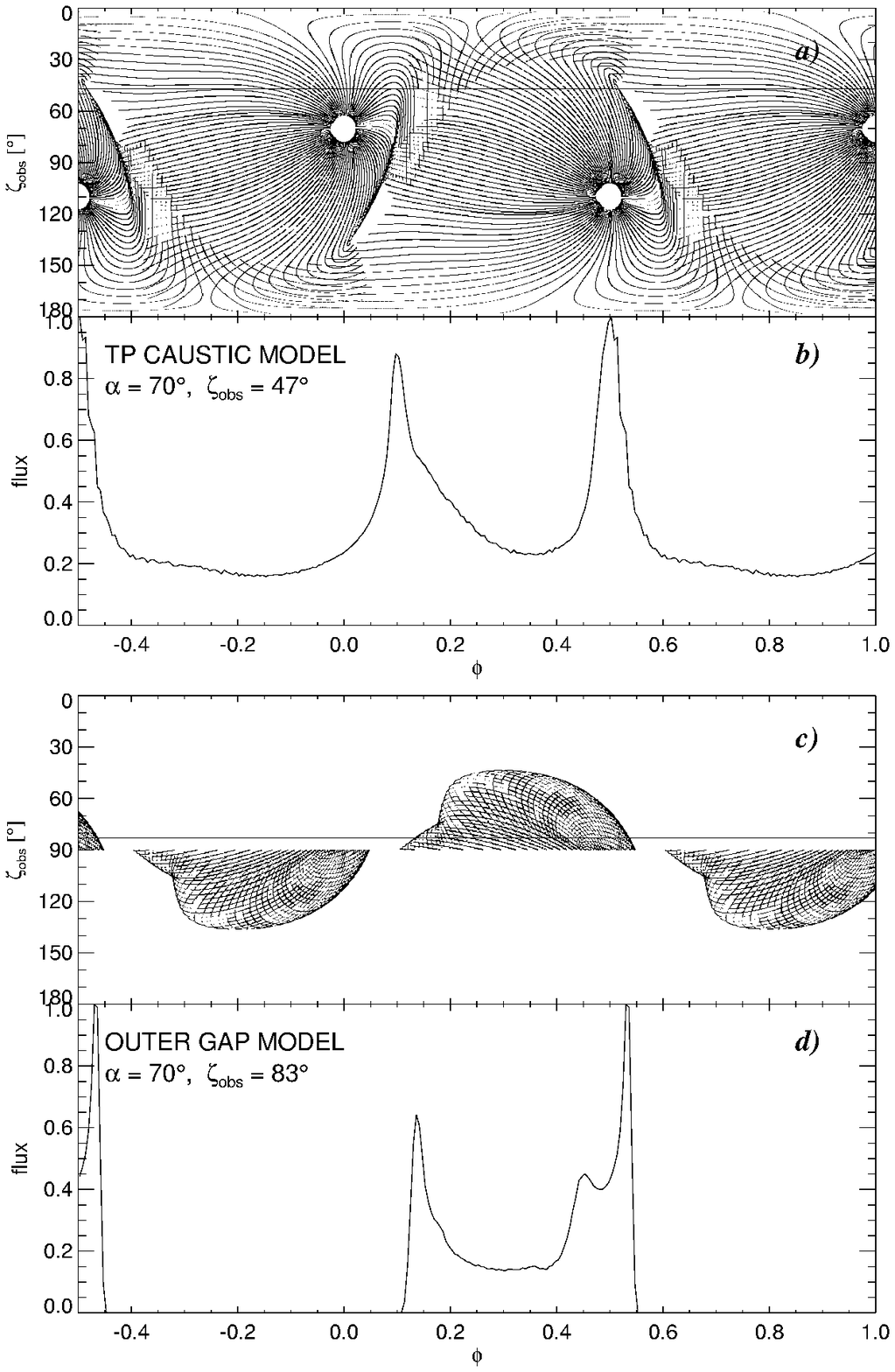}
\caption{\bm $\bf \it a)$\ubm A pattern of pulsar high-energy emission
on the $\phizet$ plane calculated with the two-pole caustic
model for $\alpha=70^\circ$, $\rovc=1$ (last open $B$-field lines), 
$\rmax=\rlc$, $\rhomax=0.8\rlc$ and $P=0.033$ s.
$\phi$ is the rotational phase, and $\zobs$ is the viewing angle
measured from the rotation axis.
\bm $\bf \it b)$\ubm A high-energy lightcurve predicted by the TPC model
for an observer located at $\zobs=47^\circ$, ie.~a 
horizontal crossection of the pattern from panel {\it a}) at $\zobs$ marked
with the horizontal line. The transverse gaussian emissivity profile
with $\sigma=0.025$, centered at $\rovc=1$ was assumed.
\bm $\bf \it c)$ and $\bf \it d)$\ubm -- the same as {\it a}) and {\it b}) respectively, but for the
outer gap model with $\zobs=83^\circ$, $\rovc=0.85$, $\rmax=1.7\rlc$ and
$\rhomax=0.999\rlc$. 
Note prominent differences
between the models.
}
\label{f2}
\end{figure}

\begin{figure}[!t]
\includegraphics[width=\textwidth]{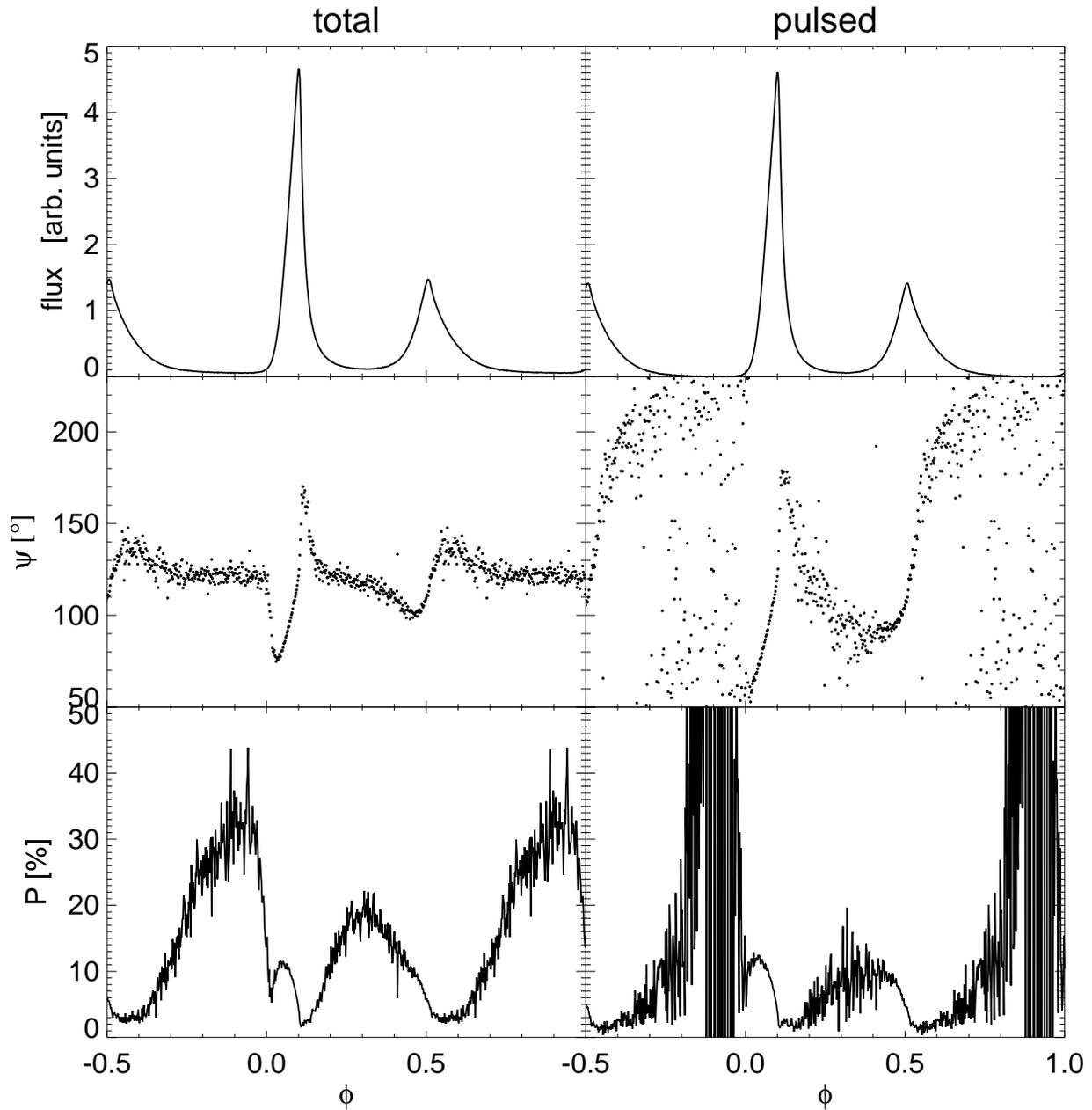}
\caption{Preliminary optical data on the Crab pulsar
obtained with the OPTIMA instrument (Kanbach et al.~2003; this workshop).
{\bf Left column:} a lightcurve (top panel), a position angle $\psi$ (dots,
middle panel),
and a degree of polarization $P[\%]$ (bottom panel). 
The constant value of position angle within phase ranges $0.2 - 0.3$
and $0.7-1.0$ suggests that the received radiation consists of two
components, one of which has constant properties.
{\bf Right column:} same as in the left column but with the contribution
of
the constant component subtracted from the data. Following Kellner 
(2002), for the constant component we assumed intensity equal to $1.24$
\% of the maximum intensity of the total signal, $\psi=123^\circ$,
and $P=33$ \%.
One and a half period is shown. The maximum of the leading peak
was aligned with the phase $\phi=0.1$.
The data were kindly provided by G.~Kanbach.
}
\label{f3}
\end{figure}

\begin{figure}[!t]
\includegraphics[width=\textwidth]{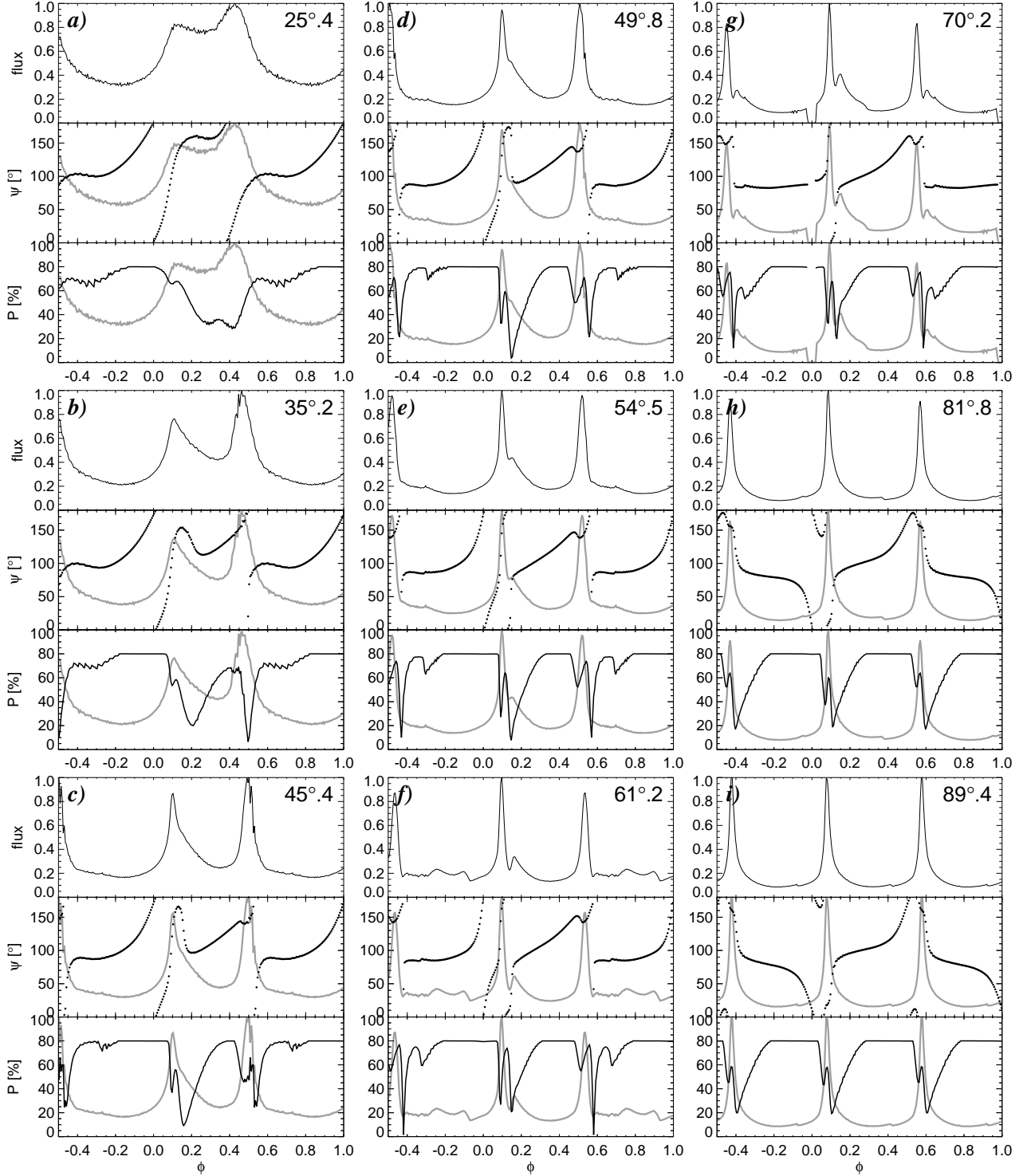}
\caption{Radiation characteristics predicted by the {\bf two-pole
caustic}
model for a pulsar with dipole inclination $\alpha = 70^\circ$.
Nine three-panel frames correspond to nine different viewing angles
$\zobs$ (marked in the top right corners).
Each frame presents the lightcurve (top panel), the
position angle curve (dots, middle panel), and the degree of linear
polarization (thick solid line, bottom panel). For reference,
the lightcurve is overplotted in the middle and in the bottom panels
as a thick grey line.
Note the dominance of two widely separated peaks in lightcurves
for most viewing angles, as well as the fast swings
of the position angle and minima in polarization degree at/close to the
peaks.
The results are for photon emission constrained to $\rho\le0.8\rlc$ 
and $r\le \rlc$.
}
\label{f4}
\end{figure}

\begin{figure}[!t]
\includegraphics[width=\textwidth]{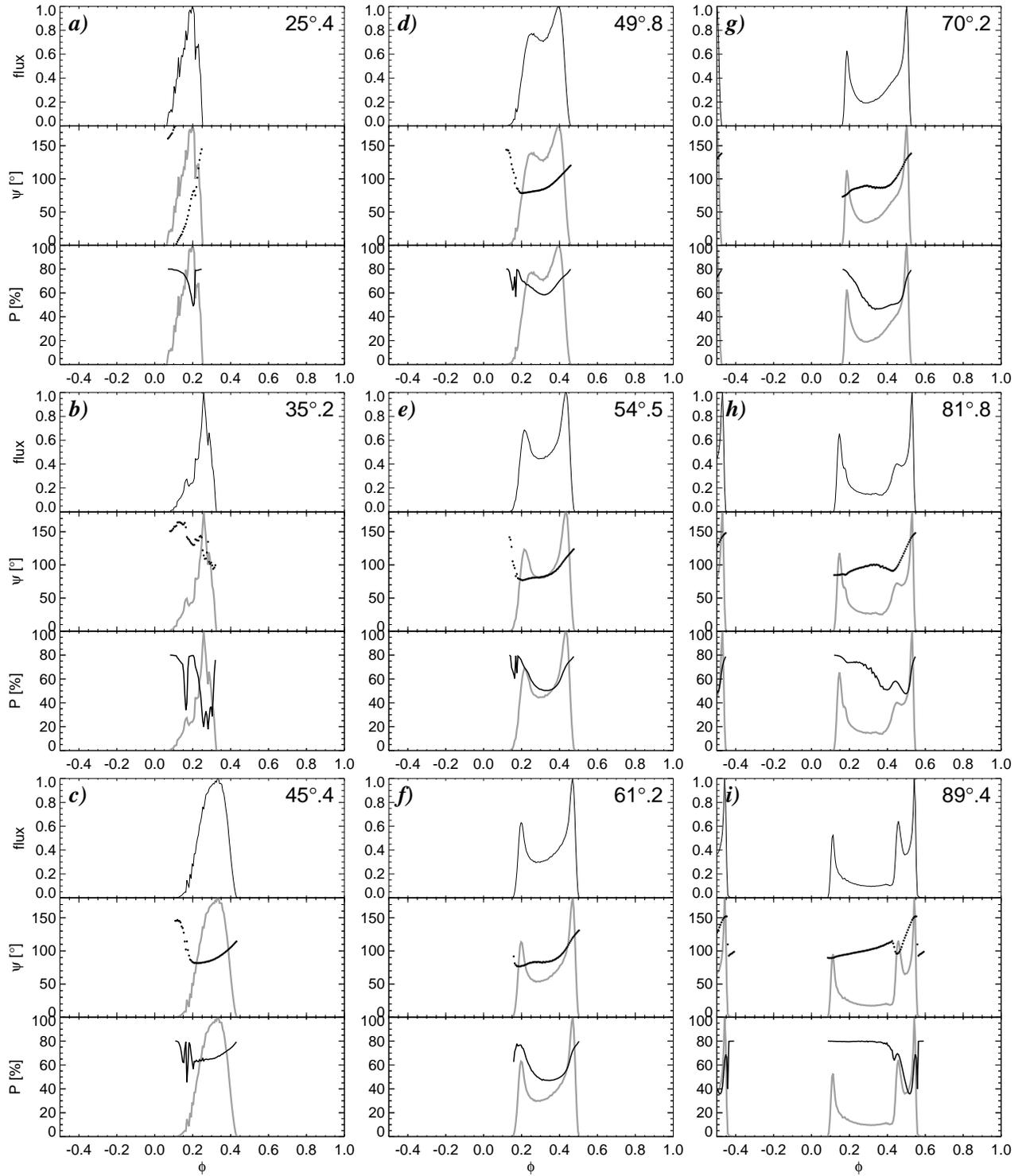}
\caption{Radiation characteristics predicted by the {\bf outer gap}
model for a pulsar with dipole inclination $\alpha = 70^\circ$.
The layout is the same as in Fig.~4.
Emission region was constrained to $\rho\le0.999\rlc$ 
and $r\le 1.7\rlc$.
}
\label{f5}
\end{figure}

\begin{figure}[!t]
\centering
\includegraphics[width=0.6\textwidth]{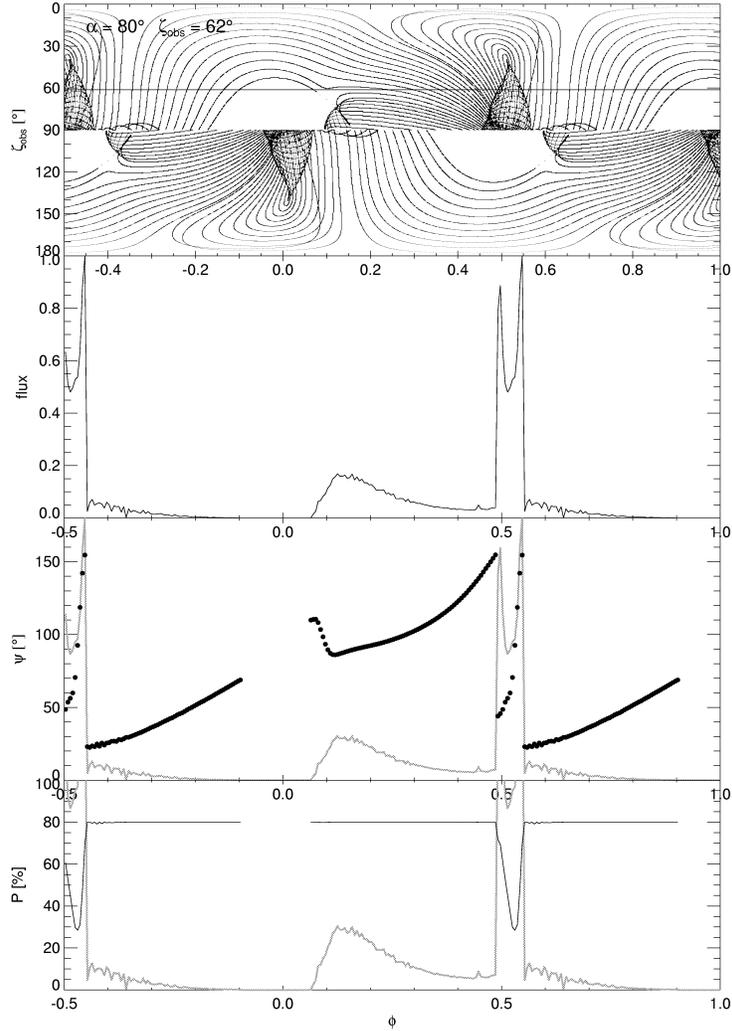}
\caption{Top to bottom: the radiation pattern, the lightcurve, the
position angle curve, and the degree of polarization predicted by the
{\bf outer gap} model for the same parameters as in fig.~5 of RY95 ($\alpha =
80^\circ$, $\zobs=62^\circ$). The emissivity along magnetic field
lines as well as the choice of these gamma-ray-bright lines was assumed
after RY95. Footprints of some of these lines on the polar cap are shown
in the next figure.  
}
\label{roger}
\end{figure}

\begin{figure}[!t]
\centering
\includegraphics[width=0.7\textwidth]{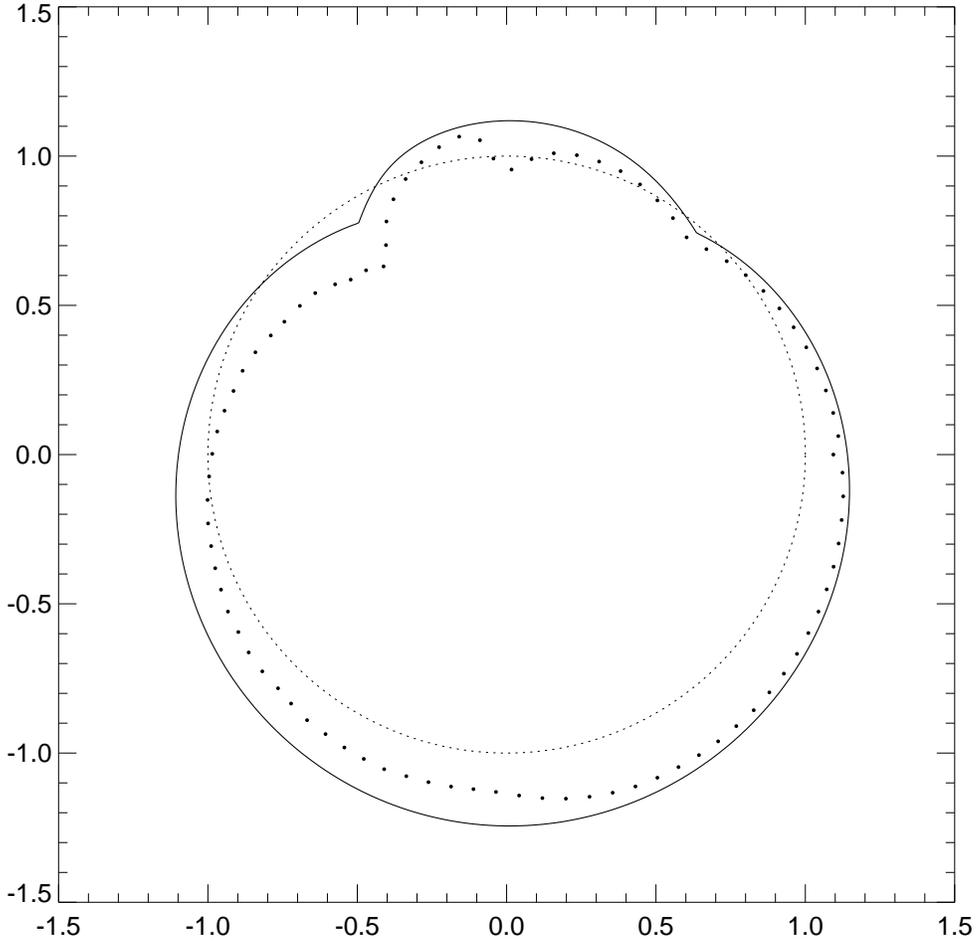}
\caption{Footprints of the gamma-ray-bright magnetic field lines
on the polar cap surface (sparse dots), calculated for $\alpha=80^\circ$
according to the formula
$w=0.02(70^\circ/\alpha)(r_c/r_{c{\rm , min}})$ from RY95.
The thick solid line is the rim of the polar cap, and the dotted line is
a circle of the standard polar cap radius $r_{\rm pc} = (R_{\rm NS}^3/\rlc)^{1/2}$. To reproduce the results of RY95, the footprints on the ``equatorward"
side of the polar cap (on the right-hand side in the figure) would have
to lie farther apart from the rim of the polar cap.
}
\label{inelpos}
\end{figure}
\end{document}